\documentclass[english]{article}
\usepackage[T1]{fontenc}
\usepackage[latin9]{inputenc}
\usepackage{graphicx}

\makeatletter
\@ifundefined{date}{}{\date{}}
\makeatother

\usepackage{babel}
\begin{document}
\title{Towards Weak Source Coding}
\author{Aman Chawla}
\maketitle
\begin{abstract}
In this paper, the authors provide a weak decoding version of the
traditional source coding theorem of Claude Shannon. The central bound
that is obtained is {\normalsize{}
\[
\chi>\log_{\epsilon}(2^{-n(H(X)+\epsilon)})
\]
where 
\[
\chi=\frac{\log(k)}{n(H(X)+\epsilon)}
\]
and $k$ is the number of unsupervised learning classes formed out
of the non-typical source sequences. The bound leads to the conclusion
that if the number of classes is high enough, the reliability function
might possibly be improved. The specific regime in which this improvement
might be allowable is the one in which the atypical-sequence clusters
are small in size and sparsely placed; similar regimes might also
show an improvement. }{\normalsize\par}
\end{abstract}

\section{Introduction}

In this paper we will attempt to lower the upper bound $\epsilon$
on the source coding probability of error. The paper assumes readers'
familiarity with source coding. Consider that in traditional source
coding\footnote{A very simply described - and hence accessible - version of Shannon's
source coding theorem can be found in \cite{wikipediaSourceCoding}.}, the probability of error of the encoder is bounded above by a small
positive constant $\epsilon$. This means that the error-exponent,
which captures the rate of decay of the probability of error with
block length, is bounded below by $-\frac{\ln\epsilon}{N}$. So if
we change $\epsilon\rightarrow\frac{\epsilon}{2}$, say, and so reduce
the upper bound, it is equivalent to adding $\frac{\ln2}{N}$ to the
lower bound. A larger lower bound means that the rate of decay of
the probability of error with blocklength is higher. In this work,
we will thus aim for a reduction in the upper bound.

In the traditional source coding procedure, if the input sequence
doesn't lie in the typical set, the encoder outputs an arbitrary $n(H(X)+\epsilon)$
bit number which represents the error condition. In this paper we
will ask the question whether we can do better in a specific sense.

\begin{figure}
\caption{The space of all sequences is divided into the typical and atypical
sets.\label{fig:The-space-of}}

\begin{centering}
\includegraphics[width=4in]{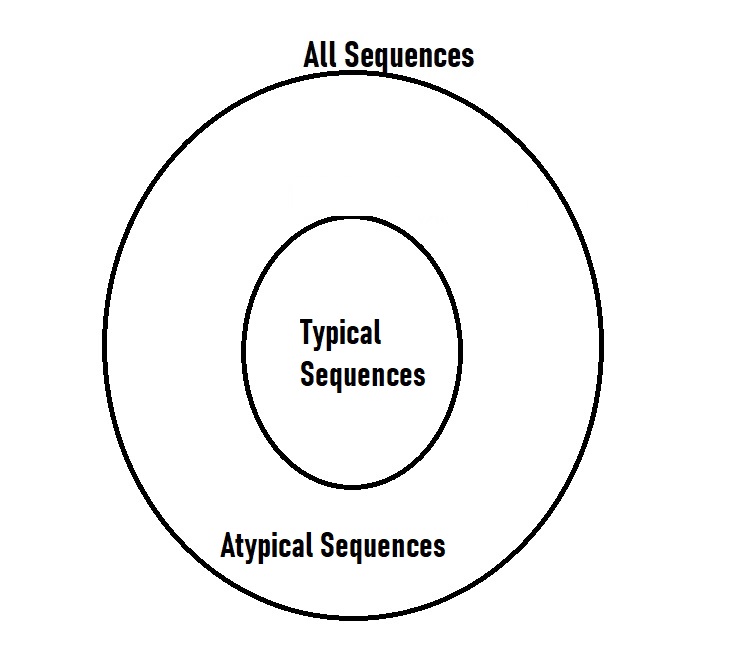}
\par\end{centering}
\end{figure}

Figure \ref{fig:The-space-of} shows the space of all sequences divided
into the typical and atypical subsets. The typical sequences are encoded
according to their position in a numbered list. They constitute a
high probability set. The atypical sequences lead to the error declaration.

\section{Innovation and its Analysis\label{sec:Analysis-of-Innovation}}

Let $A_{\epsilon}^{n}$ denote the typical set and let $B_{\epsilon}^{n}$
be the associated non-typical set. Their union is the space of all
sequences, $\mathcal{X}^{n}$. If the sequence $x^{n}$ does not belong
to the typical set, instead of declaring an error, we classify the
sequence into one of $k$ classes or clusters within $B_{\epsilon}^{n}$.
Suppose we classify it into the $j^{th}$ class where $j\epsilon\left\{ 1,...,k\right\} $.
Further, suppose the classification obeys the rule that $d(x^{n},\mu_{j})$
is the lowest amongst choices of $j\epsilon\left\{ 1,...,k\right\} $,
$\mu$ standing for the mean of a cluster.

In more detail, we output an error, but specify that it is of the
$j^{th}$ type. Since we want to encode the source, we attempt to
do unsupervised learning within $B_{\epsilon}^{n}$. We keep building
up $k$ clusters whenever $x^{n}$ lies in $B_{\epsilon}^{n}$. After
a while of `training,' the cluster-formation will more or less represent
the part of the source that generates $B_{\epsilon}^{n}$ members.

We will extend the encoding scheme in such a way that $\log(k)$ bits
will be appended to the $n(H(X)+\epsilon)$ bits previously created
for the typical sequences. If we include another bit to specify that
we are now beginning to index from the atypical set, then the net
length of the encoding becomes $1+n(H(X)+\epsilon)$+$\log(k)$ bits. 

At the same time, the error will be constrained by the size of the
largest of the $k$ clusters. Since $k\geq2$, the size of the largest
cluster belongs to the closed interval $[1,|B_{\epsilon}^{n}|-k+1]$.
In other words, the ambiguity is constrained by $|B_{\epsilon}^{n}|-k+1$
as opposed to $|B_{\epsilon}^{n}|$. These $(k-1)$ fewer sequences
are the means of the unity-sized (non-largest) clusters. Let $B'$
denote the set with the size $|B_{\epsilon}^{n}|-k+1$. Since $B_{\epsilon}^{n}$
is the atypical set, the probability of lying in it is upper-bounded
(strictly) by $\epsilon$. In this section, we are interested in saying
something about the \emph{probability of lying in $B'$}. 

Note that $B_{\epsilon}^{n}$ contains such sequences $x^{n}$ as
having $p(x_{1},...,x_{n})<2^{-n(H(X)+\epsilon)}$ or $p(x_{1},...,x_{n})>2^{-n(H(X)-\epsilon)}$.
Denote the former inequality as specifying a VLPZ or Very Low Probability
Zone and the latter inequality as specifying a VHPZ or Very High Probability
Zone. The typical sequences lie in MPZ or the Medium Probability Zone
between these two extremes. Also note that sequences within a cluster
are ``close'' to each other, i.e., they are separated by a few bit
flips only. Thus, their probabilities are alike. Hence, a cluster will
lie either completely in VHPZ or completely in VLPZ.

Suppose the largest cluster lies in VLPZ and the other $k-1$ mean
sequences are in VHPZ. Then the probability of $B'$ is approximately
$0$. On the other hand, suppose the largest cluster lies in VHPZ
and the other $k-1$ mean sequences are in VLPZ. Then the probability
of $B'$ is the probability of the largest cluster. Denote this probability
as PLC. Clearly PLC is strictly less than the probability of $B_{\epsilon}^{n}$
which in turn is strictly less than $\epsilon$.

\section{Conclusion\label{sec:Conclusion}}

To summarize, if PoE1 is strictly less than $\epsilon$, and requires
$n(H(X)+\epsilon)$ bits for operationalization\footnote{By operationalization, we refer to the process of enumerating the
members of the set whose probability we are discussing. This enumeration
requires a certain number of bits for the description of the enumerated
sequences.}, and additionally, if PoE2 is equal to PLC which is strictly less
than PoE1, and requires $1+n(H(X)+\epsilon)+\log(k)$ bits for operationalization,
it is easy to show, as is done in the Appendix,
that the exponent corresponding to the second situation can possibly
be better in some similar settings.

To conclude, we have demonstrated that there may be room for improvement
in the traditional source coding error exponent by making use of a
simple $k$-means clustering approach to the atypical sequences. Thus
machine learning might be of aid in generating (marginal) improvements
to the source coding reliability function. In upcoming work, the authors
will focus on a quantum information theoretic version of the present
paper. We will also utilize MATLAB simulations to concretely see the
gains from using machine learning, as was done in the channel coding
case \cite{chawlamorgeraweak12022}.

\section*{Appendix 1\label{sec:Appendix-1}}

Consider two cases: 

\subsection*{Case 1}

The error exponent is $e_{1}$. The PoE is PoE1. The blocklength of
sequences is $N_{1}$. Thus, we can write:

\begin{equation}
e_{1}=-\frac{\ln(PoE1)}{N_{1}}
\end{equation}

\subsection*{Case 2}

The error exponent is $e_{2}$. The PoE is PoE2. The blocklength of
sequences is $N_{2}$. Thus, we can write: 

\begin{equation}
e_{2}=-\frac{\ln(PoE2)}{N_{2}}
\end{equation}

From Section \ref{sec:Conclusion}, we have 

\begin{equation}
PoE1<\epsilon,
\end{equation}

\begin{equation}
N_{1}=n(H(X)+\epsilon),
\end{equation}

\begin{equation}
PoE2=PLC
\end{equation}
and 

\begin{equation}
N_{2}=1+n(H(X)+\epsilon)+\log(k).
\end{equation}
We ask the question: when is $e_{2}>e_{1}$? Expanding the LHS and
the RHS in this question, we obtain the relation 

\begin{equation}
-\frac{\ln(PoE2)}{N_{2}}>-\frac{\ln(PoE1)}{N_{1}}
\end{equation}
from the definitions. This implies that 

\begin{equation}
-\frac{\ln(PLC)}{1+n(H(X)+\epsilon)+\log(k)}>-\frac{\ln(PoE1)}{n(H(X)+\epsilon)}.
\end{equation}
Upon multiplying throughout by negative unity, we get

\begin{equation}
\frac{\ln(PLC)}{1+n(H(X)+\epsilon)+\log(k)}<\frac{\ln(PoE1)}{n(H(X)+\epsilon)}.
\end{equation}
Utilizing the bound on $PoE1$ we get

\begin{equation}
\frac{\ln(PLC)}{1+n(H(X)+\epsilon)+\log(k)}<\frac{\ln(\epsilon)}{n(H(X)+\epsilon)}.
\end{equation}
Upon rearranging we get

\begin{equation}
\ln(PLC)<\frac{\ln(\epsilon)\cdot(1+n(H(X)+\epsilon)+\log(k))}{n(H(X)+\epsilon)}.
\end{equation}
This can be simplified to

\begin{equation}
\ln(PLC)<\ln(\epsilon)\cdot(1+\frac{1}{n(H(X)+\epsilon)}+\frac{\log(k)}{n(H(X)+\epsilon)}).
\end{equation}
For large $n$, the second term on the RHS vanishes, but the third
term can linger if the number of clusters formed within the atypical
set also increases. Denote,

\begin{equation}
\chi=\frac{\log(k)}{n(H(X)+\epsilon)}.
\end{equation}
Thus, for large $n$,

\begin{equation}
\ln(PLC)<\ln(\epsilon)\cdot(1+\chi).
\end{equation}
where $\chi>0$. These considerations imply that $e_{2}<e_{1}.$ However,
note that this is only one setting and was described in Section \ref{sec:Analysis-of-Innovation}.

We next study the other extreme case wherein there were exactly $k$
sequences in $B_{\epsilon}^{n}$, each forming its own cluster. Here
suppose that the largest cluster is in VLPZ. Then $P(B')<2^{-n(H(X)+\epsilon)}$.
On the other hand, if the largest cluster is in VHPZ, then $P(B')>2^{-n(H(X)-\epsilon)}$.
Since the error depends upon the size of the largest cluster, we can
quantify it exactly as $PoE2=P(\mathrm{\mathrm{Cluster}1mean})\cdot P(\mathrm{Cluster1})+...+P(\mathrm{Clusterkmean})\cdot P(\mathrm{Clusterk})$.
This gives the upper bound $PoE2<\epsilon\cdot2^{-n(H(X)+\epsilon)}$. 

We use the just obtained bound. We have 

\begin{equation}
PoE1<\epsilon,\label{eq:poe1bound}
\end{equation}

\begin{equation}
N_{1}=n(H(X)+\epsilon),
\end{equation}

\begin{equation}
PoE2<\epsilon\cdot2^{-n(H(X)+\epsilon)}\label{eq:poe2bound}
\end{equation}
and 

\begin{equation}
N_{2}=1+n(H(X)+\epsilon)+\log(k).
\end{equation}
We again ask the question: when is $e_{2}>e_{1}$? Expanding the LHS
and the RHS in this question, we obtain the relation 

\begin{equation}
-\frac{\ln(PoE2)}{N_{2}}>-\frac{\ln(PoE1)}{N_{1}}
\end{equation}
which implies that 

\begin{equation}
-\frac{\ln(PoE2)}{1+n(H(X)+\epsilon)+\log(k)}>-\frac{\ln(PoE1)}{n(H(X)+\epsilon)}.
\end{equation}
Upon multiplying throughout by negative unity we get

\begin{equation}
\frac{\ln(PoE2)}{1+n(H(X)+\epsilon)+\log(k)}<\frac{\ln(PoE1)}{n(H(X)+\epsilon)}.
\end{equation}
Utilizing the bound (Equation (\ref{eq:poe1bound})) on $PoE1$ we
get

\begin{equation}
\frac{\ln(PoE2)}{1+n(H(X)+\epsilon)+\log(k)}<\frac{\ln(\epsilon)}{n(H(X)+\epsilon)}.
\end{equation}
Upon rearranging we get

\begin{equation}
\ln(PoE2)<\frac{\ln(\epsilon)\cdot(1+n(H(X)+\epsilon)+\log(k))}{n(H(X)+\epsilon)}.
\end{equation}
This can be simplified to

\begin{equation}
\ln(PoE2)<\ln(\epsilon)\cdot(1+\frac{1}{n(H(X)+\epsilon)}+\frac{\log(k)}{n(H(X)+\epsilon)}).
\end{equation}
For large $n$, the second term on the RHS vanishes, but the third
term can linger if the number of clusters formed within the atypical
set also increases. Denote,

\begin{equation}
\chi=\frac{\log(k)}{n(H(X)+\epsilon)}.
\end{equation}
Thus, for large $n$,

\begin{equation}
\ln(PoE2)<\ln(\epsilon)\cdot(1+\chi).\label{eq:poe2secondextreme}
\end{equation}
where $\chi>0$. Compare Equation (\ref{eq:poe2secondextreme}) with, 

\begin{equation}
\ln(PoE2)<\ln(\epsilon)+\ln(2^{-n(H(X)+\epsilon)})
\end{equation}
which is the inequality one obtains based on Equation (\ref{eq:poe2bound}).
The comparison yields that both inequalities can be satisfied if

\begin{equation}
\chi>\log_{\epsilon}(2^{-n(H(X)+\epsilon)}),
\end{equation}
which, being a non-integer-logarithmic lower bound that involves the
number of clusters $k$ on the LHS, shows that there is a region wherein
having $k$-means clustering in the atypical set can improve the reliability
function in this extreme setting. Since the usual setting will possibly
lie somewhere in between the two extreme regimes studied in Section
\ref{sec:Analysis-of-Innovation} and the present Appendix,
we may conclude that a gain (i.e. $e_{2}>e_{1}$) from machine learning
is \emph{not} ruled out.

\bibliographystyle{unsrt}
\bibliography{mylib2}

\end{document}